%% file: main.tex
\PassOptionsToPackage{table,dvipsnames}{xcolor}
\documentclass[sigconf,screen]{acmart}

\AtBeginDocument{%
  }

\usepackage{tikz}
\usepackage{amsmath}

\usepackage{filecontents}

\usepackage{subcaption}
\usepackage{multirow}
\usepackage{tcolorbox}
\usepackage{url}

\newcommand{\squishlist}{
 \begin{list}{$\bullet$}
   { \setlength{\itemsep}{0pt}
     \setlength{\parsep}{0pt}
     \setlength{\topsep}{0pt}
     \setlength{\partopsep}{0pt}
     \setlength{\leftmargin}{2.5em}
     \setlength{\labelwidth}{1.5em}
     \setlength{\labelsep}{0.5em} } } 

\newcommand{\squishend}{
  \end{list}  }

\AtBeginDocument{%
  }

\acmConference[HotSoS '25]{Hot Topics in the Science of Security }{April 2025}{virtual}
\acmISBN{978-1-4503-XXXX-X/18/06}

\begin{document}

\title{A Human Study of Cognitive Biases in Web Application Security}

\author{Yuwei Yang}
\affiliation{%
 \institution{Vanderbilt University}
 \city{Nashville}
 \state{tennessee}
 \country{USA}
}
\email{yuwei.yang@vanderbilt.edu}
\author{Skyler Grandel}
\affiliation{%
 \institution{Vanderbilt University}
 \city{Nashville}
 \state{tennessee}
 \country{USA}
}
\email{skyler.h.grandel@vanderbilt.edu}
\author{Daniel Balasubramanian}
\affiliation{%
 \institution{Vanderbilt University}
 \city{Nashville}
 \state{tennessee}
 \country{USA}
}
\email{daniel.a.balasubramanian@vanderbilt.edu}
\author{Yu Huang}
\affiliation{%
 \institution{Vanderbilt University}
 \city{Nashville}
 \state{tennessee}
 \country{USA}
}
\email{yu.huang@vanderbilt.edu}
\author{Kevin Leach}
\affiliation{%
 \institution{Vanderbilt University}
 \city{Nashville}
 \state{tennessee}
 \country{USA}
}
\email{kevin.leach@vanderbilt.edu}


\renewcommand{\shortauthors}{}

\begin{abstract}
Cybersecurity training has become a crucial part of computer science education and industrial onboarding. 
Capture the Flag (CTF) competitions have emerged as a valuable, gamified approach for developing and refining the skills of cybersecurity and software engineering professionals. 
However, while CTFs provide a controlled environment for tackling real-world challenges, the participants' decision-making and problem-solving processes remain under explored.
Recognizing that psychology may play a role in a cyber attacker's behavior, we investigate how cognitive biases could be used to improve CTF
education and security.
In this paper, we present an approach to control cognitive biases, specifically Satisfaction of Search and Loss Aversion, to influence and potentially hinder attackers' effectiveness against web application vulnerabilities in a CTF-style challenge.

We employ a rigorous quantitative and qualitative analysis through a controlled human study of CTF tasks.
CTF exercises are widely-used in cybersecurity education and research to simulate real-world attack scenarios and help participants develop critical skills by solving security challenges in controlled environments.
In our study, participants interact with a web application containing deliberately embedded vulnerabilities while being subjected to tasks designed to trigger cognitive biases.
Our study reveals that many participants exhibit the Satisfaction of Search bias and that this bias has a significant effect on their success.
On average, participants found 25\% fewer flags compared to those who did not exhibit this bias.
Our findings provide valuable insights into how cognitive biases can be strategically employed to enhance cybersecurity outcomes, education, and measurements through the lens of CTF challenges.
\end{abstract}

\ccsdesc[500]{Security and privacy~Usability in security and privacy}
\ccsdesc[300]{Security and privacy~Social aspects of security and privacy}
\ccsdesc[300]{Applied computing~Interactive learning environments}

\keywords{Cybersecurity, Human Aspects of Software Engineering, Capture the Flag, Cognitive Biases}



\maketitle

\section{Introduction}
Computer systems continually face threats from unauthorized access attempts~\cite{6513420},
leading to costly hacking campaigns~\cite{cashell2004economic}. 
In response to these threats, it has become essential to train cybersecurity professionals to effectively defend against such attacks ~\cite{vsvabensky2021cybersecurity}.

In this context, cybersecurity education has become a critical part of academic curricula for computer science, and Capture the Flag (CTF) competitions have emerged as an engaging, gamified approach for students to practice and refine their skills~\cite{mcdaniel2016capture}.
We are particularly interested in exploring how cognitive processes, such as decision-making and problem-solving, affect CTF performance, and whether these insights can help improve participants' skills and outcomes in such competitions.

The Tularosa Study highlighted how crucial defensive deception is in changing the attacker's decision-making process, which increases a CTF player's workload and decreases a defender's~\cite{ferguson2018tularosa}. 
However, the exploration of specific cognitive biases in this context remains under explored. 

Cognitive biases are systematic patterns of deviation from normal or rational judgment, which can significantly shape decision-making processes~\cite{haselton2015evolution}. 
By leveraging cognitive biases, we hypothesize that defenders can potentially strategically alter an CTF player's perception of the system, thereby affecting their behavior and performance~\cite{barach2021satisfaction, gachter2022individual}.

In this paper, we pursue three main objectives: 
(1) designing instrumentation to evaluate CTF player's preferences and vulnerability discovery, 
(2) examining the influence of cognitive biases on how participants attempt to compromise a web application, and 
(3) assessing changes in participant affective states as a result of inducing cognitive biases in a CTF setting.
Specifically, we investigate the effects of \emph{Loss Aversion} (LA) and \emph{Satisfaction of Search} (SoS) on these individuals' actions.
These biases are well-understood in other domains, such as economics ~\cite{novemsky2005boundaries} and radiology ~\cite{berbaum1990satisfaction}, 
but their application in the realm of computer security has not been adequately explored.
Loss Aversion refers to the tendency to prefer avoiding losses over acquiring equivalent gains~\cite{novemsky2005boundaries, tversky1991loss, kahneman1984choices}.
LA has recently been cited as a significant factor influencing human decision-making in cybersecurity contexts~\cite{iarpa_rescind_baa}.
This makes it a valuable bias to exploit in cybersecurity defenses, as the fear of losing progress or rewards can potentially deter participants from continuing their efforts.

Satisfaction of Search is a common cognitive bias where individuals cease their search for solutions once a satisfactory one is found, often leading to missed or overlooked opportunities~\cite{fleck2010generalized, barach2021satisfaction}.
This bias is particularly relevant in cybersecurity, as participants may prematurely stop their attacks if they falsely believe they have achieved their goal. 
In this context, SoS could be exploited using honeypots~\cite{franco2021survey,selvaraj2016honey} to distract or measure attackers. 
By incorporating SoS into our experimental design, we aim to understand how creating a false sense of satisfaction can influence participants to abandon their efforts early.

\begin{figure*}[tb]
\centering
\includegraphics[width=\textwidth]{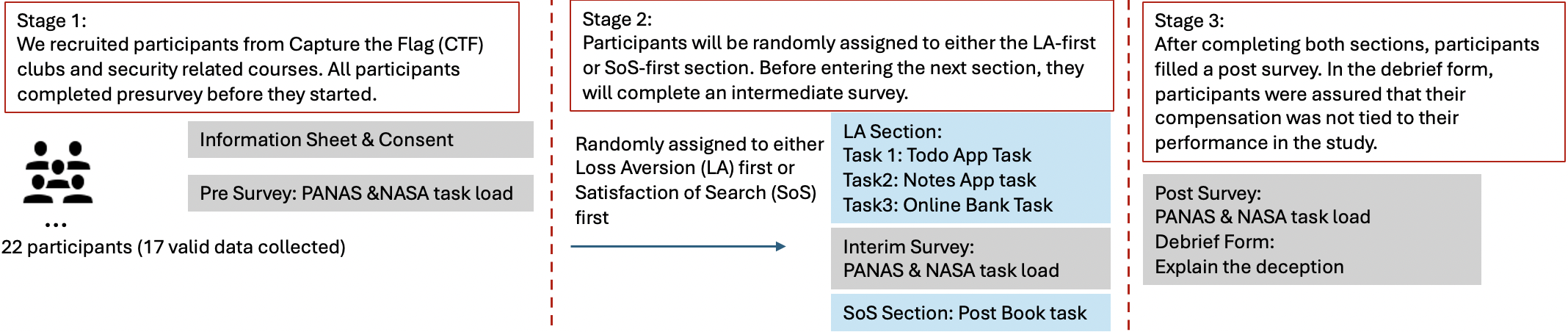}
\caption{Study Procedure. Participants first completed an information sheet and pre-survey, then were randomly assigned to either the Loss Aversion (LA) or Satisfaction of Search (SoS) task. After finishing their first task and an intermediate survey, they completed the second task. All participants finished both tasks and a post-survey, followed by a debrief form. Although participants were told that the compensation is performance based, they were all ultimately compensated \$50 regardless of performance.}
\label{fig_procedure}
\end{figure*}

To determine the manner and extent that CTF players are affected by these biases, we conducted a controlled experiment with human participants acting as attackers in an instrumented environment. 
The study procedure is shown in Figure~\ref{fig_procedure}. 
The detailed process is discussed in Study Design ~\ref{study_design}.

Through rigorous quantitative and qualitative analysis of the surveys (pre, interim, and post-surveys) and performance measurements (key logging data, number of completed tasks, and number of explored flags) from 17 participants, we observed the following key findings: 
(1) Satisfaction of Search significantly decreases participants' attacking performance; 
(2) Satisfaction of Search notably alters participants' emotional states; 
(3) Loss Aversion does not significantly impact participants' decisions to continue with security decision-making tasks; 
(4) Participants tend to explore the same type of vulnerability repeatedly.

The main contributions of this paper are as follows: 
\squishlist

    \item An IRB-approved controlled experiment to investigate the impact of cognitive biases on human participants. 

    \item A framework that integrates cognitive biases into Capture the Flag tasks, simulating realistic scenarios relevant to web security.

    \item An analysis of cognitive biases' influence on participants' performance and affective status. 

    \item An investigation of participants' cognitive patterns and exploring preferences when attacking a web application. 

    \item A discussion and set of suggestions for the implications of cognitive biases on CTF education and cyber defense.
    
\squishend

\section{Background and Related Work}
In the field of cybersecurity, understanding and mitigating the vulnerabilities of web applications is crucial due to their widespread use and importance in modern information systems~\cite{li2011survey}.

\subsection{Vulnerabilities}
\label{sec:vulns}
In this paper, we subject human participants to a set of web applications with seeded vulnerabilities in an instrumented environment modulated by cognitive biases.
To create a realistic and ecologically valid setting for our study, we included several common and severe web application vulnerabilities---SQL injection, Insecure Direct Object References (IDOR), and Cross-Site Scripting (XSS).


\noindent\textbf{SQL injection}. SQL injection is an attack that involves inserting or appending SQL code into input parameters, which are subsequently processed by a back-end SQL server. 
Web applications susceptible to SQL injection attacks can potentially allow an attacker to gain full access to their underlying databases and retrieve sensitive information~\cite{halfond2006classification}. 
We use SQL vulnerabilities in our experimental design due to their commonality and inherent harm~\cite{crespo2023sql}. 

\noindent\textbf{Insecure direct object references (IDOR)}.
The decision to integrate Insecure Direct Object References (IDOR) into our experimental task is grounded in its prevalence and real-world significance, exemplified by its inclusion in the Open Web Application Security Project (OWASP) Top 10 list of web application vulnerabilities~\cite{OWASP2017}. 
This vulnerability allows unauthorized individuals to access restricted resources~\cite{10276919}. 

\noindent\textbf{XSS injection}.
We include Cross-Site Scripting (XSS) vulnerabilities in our experimental design due to their prevalence, inherent risks, and the limitations of existing mitigation techniques~\cite{johari2012survey}. 
XSS vulnerabilities persist as a formidable threat in web applications, and conventional security measures often fall short in providing foolproof protection~\cite{gupta2017cross}.

\subsection{Cognitive Biases}
\label{biases}

In designing our experiment, we integrated cognitive biases to explore their influence on decision-making processes within cybersecurity contexts.
  
\noindent\textbf{Loss Aversion (LA)}.
Loss Aversion indicates that the value function is steeper for losses than for gains, meaning the psychological impact of losing a sum of money is greater than the pleasure derived from gaining the same amount~\cite{schmidt2002experimental}.

Psychological studies have designed experiments to measure and test Loss Aversion on an individual's decision making behavior given lotteries with varying odds~\cite{schmidt2002experimental}.
However, there is limited research discussing Loss Aversion's impact on attackers in the domain of cyber psychology.
Drawing inspiration from psychological studies on Loss Aversion, our experimental design incorporates elements of gambling, turning, and changing gain and loss to study participant behavior systematically.
Insights from research by Schmidt et al., Tom et al., and Sokol-Hessner et al.  guide our design, emphasizing the impact of Loss Aversion on decision-making under risk~\cite{schmidt2002experimental,sokol2019psychological,tom2007neural}.

\noindent\textbf{Satisfaction of Search (SoS)}.
Satisfaction of Search originates from radiology, in which a specific target is more likely to be missed during a radiological examination when accompanied by an additional abnormality, compared to when it is the only target present~\cite{fleck2010generalized}.

However, there is limited study explored how attackers in cybersecurity exhibit this cognitive bias. 
Our experimental design builds on Fleck et al., which considers diverse factors influencing SoS, including the relative frequency of different target types, external pressures (reward and time), and expectations about the number of targets present~\cite{fleck2010generalized}. 
In this paper, we expose participants to a task in which they could potentially find multiple potential targets (i.e., flags).  

\subsection{Measurements and Surveys}
\label{sec:measure}
In this study, participants are presented with vulnerable web applications and opportunities to continue or quit, enabling measurement of their perception of risk and performance during the tasks. 
In addition to these data points, we are further interested in participants' emotional changes and their self-evaluated success in response to these tasks.
Thus, we employed two key instruments to gauge the psychological states of our participants: the Positive and Negative Affect Schedule (PANAS) survey to evaluate their affective states and the NASA Task Load Index (NASA TLX) to measure mental workload and self evaluated performance while participants are completing the tasks.

\noindent\textbf{Positive and Negative Affect Schedule (PANAS)}.
To discern the emotional states of participants, we used PANAS.
This survey, developed by Watson, Clark, and Tellegen (1988), assesses two primary dimensions of affect: positive affect (PA) and negative affect (NA).
Participants indicate the extent to which they are currently experiencing a range of positive and negative emotions~\cite{thompson2007development}.
By employing PANAS, we aim to explore the potential relationship between participants' emotional states and their engagement with web application vulnerabilities, similar to previous works~\cite{baltaci2016stress}.

\noindent\textbf{NASA Task Load Index (NASA TLX)}.
We employed NASA-TLX to assess participants' perceived workload across various dimensions including mental, physical, and temporal demands, as well as performance, effort, and frustration levels~\cite{hart1988development}.
By using NASA-TLX, we obtained a holistic view of the cognitive and physical demands imposed by our tasks on participants, as well as their own perceptions of their performance and the effort they exerted~\cite{hart2006nasa}.
This self-assessment is crucial for interpreting how different task characteristics influence overall workload, thereby allowing us to better understand the human factors that determine task efficiency and effectiveness.

\subsection{Capture the Flag}
We deployed Capture the Flag (CTF) style web application tasks for our participants.
CTF tasks are widely recognized as an effective tool for cybersecurity education~\cite{mcdaniel2016capture, vsvabensky2021cybersecurity}.
CTF competitions simulate real-world hacking scenarios, providing participants with hands-on experience in identifying and exploiting vulnerabilities~\cite{erola2021control}
These tasks are designed to mimic the challenges faced by security professionals, making them a valuable method for developing practical skills in a controlled and ethical environment.

\section{Study Design}
\label{study_design}

In this paper, we present a human subject study of cognitive biases and the role they play in CTF player's behavior.
We designed an instrumented Capture the Flag (CTF) environment that participants would engage with to measure how CTF player exhibit cognitive biases in cybersecurity contexts, particularly those related to web applications.

    

\subsection{Participant Experience}

We recruited 22 subjects from university students with experience in CTF events or web application security for this IRB-approved study.

Some participants were fully engaged while others dropped out or did not complete all the requirements.
Three participants failed to complete the surveys, one participant chose to withdraw their data, and one participant completed the task with very low performance, indicating insufficient effort.
Consequently, we analyze the complete experimental session data for 17 participants.


\subsection{Ethical Considerations}
The study is divided into two key areas corresponding to our two selected cognitive biases: Loss Aversion (LA) and Satisfaction of Search (SoS).
To minimize the interaction between these biases, we randomly assigned the sequence of these sections to each participant.
Participants received an ID and password as credentials to gain access to the experiment platform, and all participant data was anonymized for participant privacy and safety. The study protocol they followed was approved by the Institutional Review Board (IRB).
All participants received \$50 USD in compensation upon completion of the experiment. 
However, critically, during the experiment, they were told that they earn compensation based on their performance in the tasks.
This intentional deception was a key aspect of providing participants with a sense of pressure, risk, and reward as part of modulating the effects of Loss Aversion and Satisfaction of Search.  
Nonetheless, all participants received the same \$50 amount after completing their experimental session.
They further received a debriefing form after the study to explain the deception.


\subsection{Protocol Implementation}

We designed our experimental stimulus within isolated Docker containers for each participant. 
This helped ensure participants were isolated from each other and provided a straightforward mechanism for adding new participants and recording their data in isolation.
Each participant was provided with a unique URL for their participation, which in turn was mapped to a specific container on our webserver. 

We used Flask within each participant's Docker container to serve the tasks during the experiment. 
The participant interacts with the web page to find and exploit vulnerabilities.  
We use JavaScript to record keystrokes and mouse position and events as the participant interacts with the stimulus interface during their scheduled experiment session. 

\subsection{Loss Aversion Section Task Design}
\label{sec:la-design}
Loss Aversion refers to people's tendency to prefer avoiding losses rather than acquiring equivalent gains~\cite{inesi2010power}.
Our experimental design was inspired by previous studies on LA in the field of psychology.
For example, in the experiment conducted by Tom et al., participants decided whether to accept or reject gambles that offered a 50/50 chance of gaining or losing money~\cite{tom2007neural}.
To induce risk and mimic a gambling situation so we might observe participants' decision making behavior, we carefully chose three challenges.
Each of these challenges represents a widely-used and common cybersecurity vulnerability: SQL Injection, Insecure Direct Object Reference (IDOR), and Cross-Site Scripting (XSS). These vulnerabilities and the reasons for their inclusion are described in detail in Section~\ref{sec:vulns}.
Participants are under time pressure to complete the assignment or risk losing their financial reward. 
Moreover, participants receive a warning message indicating that their attacks may be caught and investigated as they proceed, thus adding the perception of risk and realistic to the participant. 
This design is intended to simulate high-stakes nature of real-world cyber attacks where there is a risk of exposure with as the time taken increases and as the participants applies increasingly aggressive strategies. The three challenges are designed to simulate sequential steps necessary to achieve a final attack goal.

\subsubsection{Deception \& Experiment Protocol}

In the LA-modulated task, we present participants with a sequence of web applications with seeded SQL injection, IDOR, and XSS vulnerabilities.
The participant must first find the SQL injection vulnerability, then the IDOR vulnerability, and finally the XSS vulnerability. 
Participants are not told which vulnerabilities are present, but they are told they need to discover and exploit vulnerabilities.
If the participant identifies and exploits the corresponding vulnerability, they are asked whether they want to proceed to the next stage. 
At each stage, they are told they risk discovery and losing all financial gain, but that they have an opportunity to increase how much financial reward they receive if they succeed.
Thus, participants are tasked with a critical decision-making moment (i.e., whether to continue to the next stage while risking all their financial gain, or to quit and keep their current financial gain).  
During this decision, we assess participants' perception of risk and decisions under duress, allowing us to measure the impact of LA in this context. 

At each stage, the participants are asked what minimum financial return they would have accepted to take the risk to continue.
Those who choose to proceed provide important information on the scope and makeup of incentives that affect risk-taking in hacking scenarios.
On the other hand, those who choose to quit are asked the same questions, which aids in our comprehension of the barriers that prevent people from taking additional risks.
We record their decision to proceed or quit at each stage along with their keystrokes, mouse events, and psychological measurements.

\begin{figure*}[tb]
    \centering
    \includegraphics[width=\textwidth]{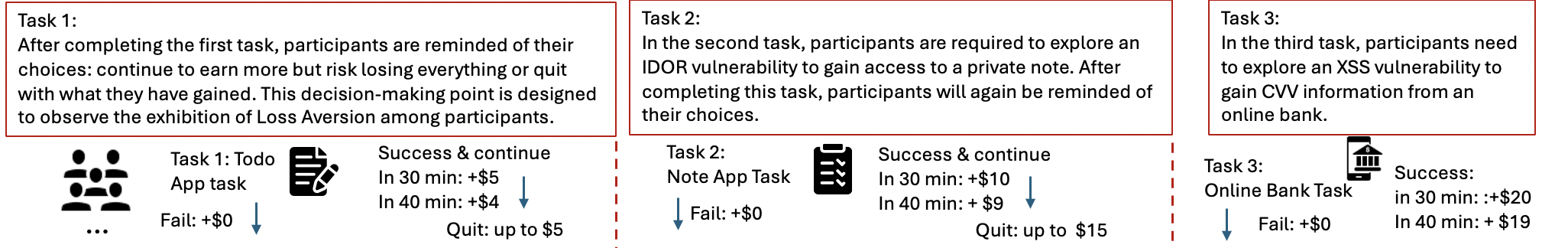}
    \caption{Participants in the Loss Aversion section had to retrieve a bank account CVV through a three-stage process.
    The first task involved using SQL injection to obtain a password from a `To-Do' application. 
    We told participants they would receive a \$5 reward for completing this task.  
    At each stage, participants chose whether to continue, with the second task offering an additional \$10 and the third \$20, but failure at any stage meant losing all accumulated rewards.
    Participants' decisions were recorded to analyze their behavior under potential loss and gain.}
    \label{fig:procedure_LA}
    \vspace{-2ex}
\end{figure*}

Our detailed experimental procedure for the LA task is illustrated in Figure~\ref{fig:procedure_LA}. 
Next, we describe the stimulus design for the three LA-modulated tasks.
We designed a separate web app for each vulnerability, which we describe below. 
\noindent\textbf{Todo App (first task): SQL Injection}.
First, we consider an SQL Injection vulnerability, a technique where CTF players manipulate standard SQL queries to gain unauthorized access to a database.
Participants are presented with a web application mimicking a 
`To-Do' app, where their objective is to uncover the password to a `Notes' app, believed to store the password for an online banking account. 
To succeed, participants must exploit SQL vulnerabilities to access the admin account of the `To-Do' app and locate the password. 
\noindent\textbf{Notes App (second task): IDOR}.
The second task focuses on IDOR, which allows unauthorized users to access to hidden resources. 
In this scenario, participants interact with the `Notes' app. 
Their goal is to find a password for online banking login. 
The task is designed such that while the first note requires a password for access, participants can bypass this by accessing it directly via the URL. 

\noindent\textbf{Online Bank (third task): XSS injection}.
The final task involves XSS, a vulnerability where CTF players inject malicious scripts into web applications. 
Participants face a web application styled as an online banking page. 
Their objective is to find the CVV number of a credit card, achievable through injecting malicious scripts via the search bar. 


\subsection{Satisfaction of Search Section Task Design}

Recall that SoS refers to the phenomenon wherein one detection of an abnormality in an image impedes the detection of additional abnormalities~\cite{ashman2000satisfaction}. 
Our experimental design is inspired by previous studies on this subject in cognitive psychology and radiology. 
For instance, in Fleck et al.'s study, researchers presented multiple visual targets to their participants in one trial and measured the accuracy of each participant's search~\cite{fleck2010generalized}. 
To execute our study, we designed a single web application concealing eight vulnerabilities --- that is, eight targets were included in one trial for each participant to find.
We hypothesize that the SoS effect will 
influence CTF players to stop searching for additional vulnerabilities after finding a small number of vulnerabilities.  
Similar to including honeypots, we investigate the impact of including seeded defects on CTF player cognition. 

As with the LA Section of our experiment (Section~\ref{sec:la-design}), we consider three prevalent vulnerabilities: 
SQL injection (SQL), 
Insecure Direct Object References (IDOR), and Cross-Site Scripting (XSS). We select these vulnerabilities due to their prevalence in web applications security (c.f. Section~\ref{sec:vulns}). 

\subsubsection{Deception and Experiment Protocol}
Participants are granted unrestricted time to identify the vulnerabilities in the given application, unaware of the total number concealed. 
Each successful identification is accompanied by 
a ``flag'' (a series of characters like a password), akin to the structure of CTF competitions. 
Participants are told that each discovery of a flag would earn them a \$2 bonus.

We designed the SoS web application to allow the participant to track how many flags they successfully acquired.
We tracked the keystrokes, mouse events, and timing associated with their interaction with the page. 
We recorded where and how they achieved each flag, which we use a basis for analyzing potential participant preferences for certain classes of vulnerabilities. 
Furthermore, our investigation extends to scrutinize participant satisfaction at varying levels of vulnerability discovery --- when the participants would be satisfied with their exploration and would stop searching for more vulnerabilities.
By discerning patterns in these behavioral aspects, we aim to extrapolate insights applicable to real-world scenarios, potentially influencing and deterring malicious hacking behavior.
Upon completing the experiment, all participants receive a debrief form, unveiling the deception.
Regardless of their choices, all participants are compensated equally, ensuring that payment does not depend on their decisions.

\subsubsection{PostBook Web Application task}

\begin{figure}[tb]     
    \centering     
    \includegraphics[width = \columnwidth]{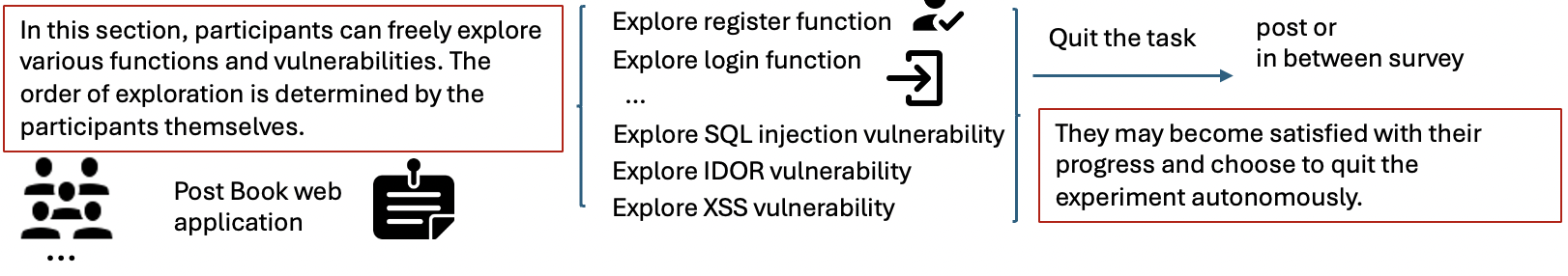}     
    \caption{Participants in the SoS experiment were tasked with exploring vulnerabilities in the PostBook web application.
There were 8 flags hidden that were accessible using three types of vulnerabilities.
During the exploration process, participants could stop at any time they wished.
No time limit was set to ensure that participants could explore the vulnerabilities at their own pace and quit when they felt satisfied.}        
    \label{fig:procedure_sos} 
    \vspace{-2ex}
\end{figure}

We designed an application that resembles a collaborative platform for posting content like a digital bulletin board~\cite{Hackerone}.
Participants engage with a spectrum of functions, including user authentication processes such as login, registration, and logout.
The detailed experimental procedure is illustrated in Figure~\ref{fig:procedure_sos}.
Upon accessing the dashboard, participants have the ability to create posts, which
harbor strategically embedded vulnerabilities.
The scope of user interaction extends to post editing, wherein participants can modify the content displayed on the dashboard.
An additional layer of complexity arises with the option to categorize posts as either public or private, affording participants the authority to control the visibility of their contributions.
Otherwise, the user can create their profile and view others' profiles.
Participants were tasked with exploring and finding vulnerabilities in this platform.
They continued to search for vulnerabilities until they felt satisfied they found as many as possible.

In this setup, we also included several default accounts, including an administrator, established and configured by a researcher prior to participant engagement. 
These default users makes the application more realistic and mimic a real world web application with existing, potentially valuable data. 

\noindent\textbf{SQL vulnerabilities}.
We plant two SQL injection vulnerabilities in the SoS task. 
First, the login function is susceptible to a straightforward SQL injection that circumvents the authentication mechanism, allowing access to arbitrary accounts on the system.
Second, the administrator account page contains several special functions that are guarded by a unique password check to access. 
Unlike the login mechanism, this second vulnerability requires a two phase injection --- first, to reset the administrator password, then a second to access the specific flag within the administrator's page.


\noindent\textbf{IDOR vulnerabilities}. 
The Insecure Direct Object References (IDOR) vulnerabilities within this experimental task are embedded in the post and profile sections of the web application, introducing a subtle yet exploitable pattern within the URL structure.
Within the post section of the web application, participants have the ability to create and edit posts. 
The IDOR vulnerability manifests through manipulation of the URL link,
which allows them to access posts authored by others, including those marked as private.
We also include a second instance of an IDOR vulnerability in the user profile portion of the stimulus web application.
This setup reflects a common IDOR vulnerability in which too little access control 
is implemented when handling GET-based HTTP requests. 



\noindent\textbf{XSS vulnerabilities}.
Participants can uncover the XSS vulnerability in this task through intentional manipulation of input strings by introducing strings that initiate malicious executable scripts. 
The vulnerability manifests when these crafted strings are submitted within the `create post' block, subsequently executing scripts under the control of the participant. 
The stimulus detects when valid JavaScript code is submitted, which in turn reveals the flag. 
We chose this design to simplify the associated attack 
--- we had to balance the time taken by participants to determine which vulnerability was present with the realism of the exploit they generated. 



\section{Data Collection and Processing}
\label{data}

This study used a comprehensive measurement technique, with a number of instruments to examine different participant behaviors and responses.
A number of questionnaires that were given out at various points during the study---a pre-survey prior to the tasks starting, an intermediate survey in between the two sections, and a post-study survey---were essential to gathering our data.
These questionnaires were intended to gather specific data regarding the respondents' prior exposure to Capture the Flag (CTF)-style tasks, their proficiency with code, and their knowledge of web application security.
In addition, the pre-survey instruments allowed discerning which participants had sufficient technical felicity with web application security and CTFs to meaningfully participate in the study. 

Aside from the self-reported surveys, our approach involved tracking participant interactions with the activities in real-time. 
We recorded keystrokes and mouse events as raw data as the participant navigated both the Loss Aversion and Satisfaction of Search tasks of our experimental protocol.


Keylogging data from the participants was meticulously labeled by a coder using a predefined codebook~\cite{macqueen1998codebook,hai2017creating}.
This qualitative approach follows best practices in existing literature.
This codebook, based on the definitions of various vulnerabilities, provided detailed guidelines to ensure consistency and accuracy in the labeling process ~\cite{halfond2006classification,pratama2022penetration,liu2019survey}.
Each keypress and mouse-click behavior was labeled with the corresponding type of vulnerability being explored.
The coder also cross-checked the labels with the participants' journals, which recorded their thoughts during the exploration process.
Our codebook can be found in our replication package (cf. Section~\ref{availability})


Recording participants' choices to either continue with the tasks or opt out at different points was an essential component of our data collection for the Loss Aversion component. 
Participants were asked to indicate the lowest incentive amount that would have affected their choice to continue or stop the tasks.
We also recorded their specific decision whether to continue or quit at each stage.

All participant data was de-identified following best practices, and we stored all data in an encrypted volume on a research lab server. 
Participants could quit at any time and receive full compensation after debriefing.

\section{Evaluation}

This section presents results and provides evaluation for the web application task and quantitative surveys. We aim to answer the following questions:
\squishlist
    \item (RQ1) Do CTF players exhibit Satisfaction of Search when exploring a system?
    \item (RQ2) Do CTF players exhibit Loss Aversion when exploring a system? 
    \item (RQ3) How do Satisfaction of Search and Loss Aversion influence the performance of participants?
    \item (RQ4) How do participants' emotional states change when they are exploring a system? 
    \item (RQ5) What cognitive patterns do participants exhibit when they are exploring a system?
\squishend



\subsection{RQ1: Exhibition of Satisfaction of Search and its Impact}
In this subsection, we analyze how Satisfaction of Search affects participants' progress and decision-making behavior, specifically measured by the number of real flags identified.

Overall, 22 participants recruited from various universities completed the task.
After removing outliers due to poor performance and incomplete survey, we retained 17 valid data points. 
Three participants failed to complete the surveys, one participant chose to withdraw their data, and one participant completed the task with very low performance, indicating insufficient effort. 
Two coders reviewed participants' responses to the question ``How did this part of the study go? Were you able to complete it?'' and their self-evaluated success. Based on these responses, the participants were grouped into two categories: those exhibiting SoS, who were satisfied with their progress and believed they had explored all vulnerabilities when they quit the study, and those who were unsure if they had found all the flags. 
The coders resolved any disagreements through discussion to finalize the groupings. Ultimately, six participants were categorized into the satisfied group. For example, one participant stated, ``I believe I completed it, and it went well,' while another mentioned, ``I'm not sure if I got all the flags, but I found a fair amount and am satisfied with my progress.'"
To measure the  inter-rater reliability among the coders, we used the Cohen's Kappa score. This resulted in a score of 0.88, which is interpreted as ``excellent'' or ``almost perfect'' agreement~\cite{fleiss2013statistical, landis1977measurement}. 

\begin{tcolorbox}[colframe=black, colback=white, boxrule=0.5mm, arc=1mm, boxsep=1mm]
\textbf{Summary:} Two coders labeled the stop reason response from the survey for Satisfaction of Search and 35.3\% of participants exhibited Satisfaction of Search (SoS) during the exploration process. 
\end{tcolorbox}



\subsection{RQ2: Exhibition of Loss Aversion}

In the Loss Aversion section of our study, participants were presented with three different tasks, each corresponding to a distinct type of vulnerability.
We established varying monetary incentives for each stage, with the stipulation that failure to complete a task would result in the loss of all previously gained rewards.
Thus, participants could complete all three tasks, run out of time (modelling the possibility of being ``caught by system administrators'' in a real world hacking attempt), or choose to quit between tasks to avoid losing their earnings (and thus exhibiting Loss Aversion).
In the end of the survey, we asked participants, ``How did the first/second part of the study go?
Were you able to complete it?'' Despite this setup, our analysis of the valid data points revealed that none of the participants chose to quit due to the fear of losing their gains.
Some participants did not complete all three tasks, but this was attributed to time constraints rather than a conscious decision to avoid risk.
Before transit to the next task, they all chose to continue rather than quit with the bonus they gain.
Some participants even consider the count down as an incentives, for example, some reflected, ``the time pressure set out by the monetary reward definitely made my mind race faster for ideas as I was looking for vulnerabilities which I believe helped me solve the tasks quicker." 

These findings suggest that Loss Aversion may not be as influential as we initially hypothesized. 
The anticipated psychological impact of potentially losing rewards did not deter participants from continuing their attempts to complete the tasks. 
This indicates that, in the context of our experiment, participants did not exhibit significant concern over losing their gains. 
They were more focused on the challenge and potential rewards than on the risk of loss.

The literature supports this observation.
According to Madarie (2017), intellectual challenge and curiosity are the strongest motivators for hackers~\cite{madarie2017hackers}.
This motivation might surpass the concerns about potential losses.
This insight is crucial for understanding the limitations of leveraging Loss Aversion as a deterrent in cybersecurity defenses.
Our results imply that CTF players, driven by these intellectual motivations, may not be easily swayed by the threat of losing rewards.
This insight is crucial for understanding the limitations of leveraging LA as a deterrent in cybersecurity defenses.
Our results imply that penetration testers, driven by other motivations, may not be easily swayed by the threat of losing rewards.
That said, we also note that our incentive of \$50 may not have been large enough to produce a substantial LA effect.
Nonetheless, while cognitive biases can still play a role in influencing CTF players behavior, the effectiveness of Loss Aversion as a standalone strategy may be limited.
Future research should explore additional factors that could be combined with Loss Aversion to create more effective deterrents in cybersecurity scenarios.

\begin{tcolorbox}[colframe=black, colback=white, boxrule=0.5mm, arc=1mm, boxsep=1mm]
\textbf{Summary:} Participants did not significantly exhibit Loss Aversion in this experiment, which might due to the motivation to explore surpassing concerns about risks.
\end{tcolorbox}

%
\subsection{RQ3: Impact from Cognitive Bias}

To explore the relationship between SoS and task performance, we employed a Linear Mixed-Effects Regression (LMER) model.
This model allowed us to account for both fixed effects (SoS) and random effects (training section order, schools, and experiment instructors).
The analysis revealed a statistically significant effect of SoS on performance, with a t-value of -2.413, estimated reduction of 2 flags, and a p-value of 0.0291.
This indicates that the presence of SoS significantly influences the number of flags identified by the participants.

The results obtained from this experiment showed that participants who exhibited SoS found fewer flags compared to those who did not.
We collect the reason why participants stop based on their responses in the post survey.
Specifically, participants experiencing SoS tended to stop searching prematurely, as they were satisfied with their progress according to their self-reported reasons for quitting and their perceived success in the task.
This premature termination of the task hindered their overall performance in identifying flags.
This finding underscores the negative impact of SoS on task performance in contexts where thoroughness and persistence are critical.

The observed relationship between SoS and reduced performance suggests that satisfaction with progress can lead to premature cessation of search activities.
This behavior is particularly effective in tasks that require exhaustive exploration and verification, like CTF events. 
Understanding the impact of SoS on decision-making and task performance can inform the design of interventions aimed at hindering the participants' exploration of the system.

\begin{tcolorbox}[colframe=black, colback=white, boxrule=0.5mm, arc=1mm, boxsep=1mm]
\textbf{Summary:} The participants exhibits Satisfaction of Search (SoS) found 25\% fewer flags on average compared to those who did not exhibit this bias.
\end{tcolorbox}

\subsection{RQ4: Affective States Changes}
In our study, we used the Positive and Negative Affect Schedule (PANAS) to assess participants' emotional states before and after engaging in specific tasks (cf. Section~\ref{sec:measure}). 
This survey provides a comprehensive measure of both positive and negative emotions, allowing us to understand the emotional impact of the tasks on participants.

To analyze the changes in emotional states, we conducted paired t-tests on the PANAS scores from our participants. These tests compared the pre- and post-task scores within each group to identify any significant changes in emotions.
In the SoS group, where participants exhibited Satisfaction of Search, we observed a significant increase in `proud' and decrease in `nervous'. The paired t-test results revealed statistically significant differences in these two emotional scores, indicating that the SoS tasks meaningfully enhanced participants' positive feelings. 
Conversely, the control group, which did not exhibit significant Satisfaction of Search, showed no significant changes in emotional scores. The paired t-tests indicated that there were no notable differences in emotions before and after the tasks for this group.

Table~\ref{table:panas} presents these findings, showing the comparison of pre- and post-task PANAS scores for both groups. The significant increase in positive emotions in the SoS group highlights the impact of SoS on enhancing participants' emotional experiences during the tasks.

\begin{table}[tb]
\centering
\caption{Results from the PANAS Form Analysis}
\vspace{-2ex}
\label{table:panas}
\begin{tabular}{lrr}
\toprule
\textbf{Emotion} & \textbf{p-value} & \textbf{Mean Difference} \\
\midrule
Excited       & 0.076 & 0.500  \\
\rowcolor{lightgray} Proud         & 0.034 & 1.167  \\
\rowcolor{lightgray} Nervous       & 0.010 & -1.333 \\
Determined    & 0.076 & 1.000  \\
Satisfied     & 0.093 & 1.667  \\
\multicolumn{3}{l}{\emph{Other emotions with $p>0.1$ are elided.}}\\
\bottomrule
\end{tabular}
\vspace{-2ex}
\end{table}

\begin{tcolorbox}[colframe=black, colback=white, boxrule=0.5mm, arc=1mm, boxsep=1mm]
\textbf{Summary:} Participants who exhibited Satisfaction of Search experienced a significant increase in feelings of pride and a decrease in feelings of nervousness. The feeling of pride may indicate overconfidence, leading participants to stop searching prematurely. 
\end{tcolorbox}

\subsection{RQ5: Cognitive Patterns}

Here we present the results of a bigram analysis conducted on the sequences of vulnerabilities explored by participants during the web application tasks.
Bigrams, or pairs of consecutive items, can provide insight into the thought patterns that participants tend to follow.
This analysis uncovered common sequences of vulnerability explorations and provided insights into participant behavior and decision-making processes.

\begin{figure}[tb]
\begin{minipage}{.48\textwidth}
    \vspace{10ex}
    \centering
        \includegraphics[width=\textwidth]{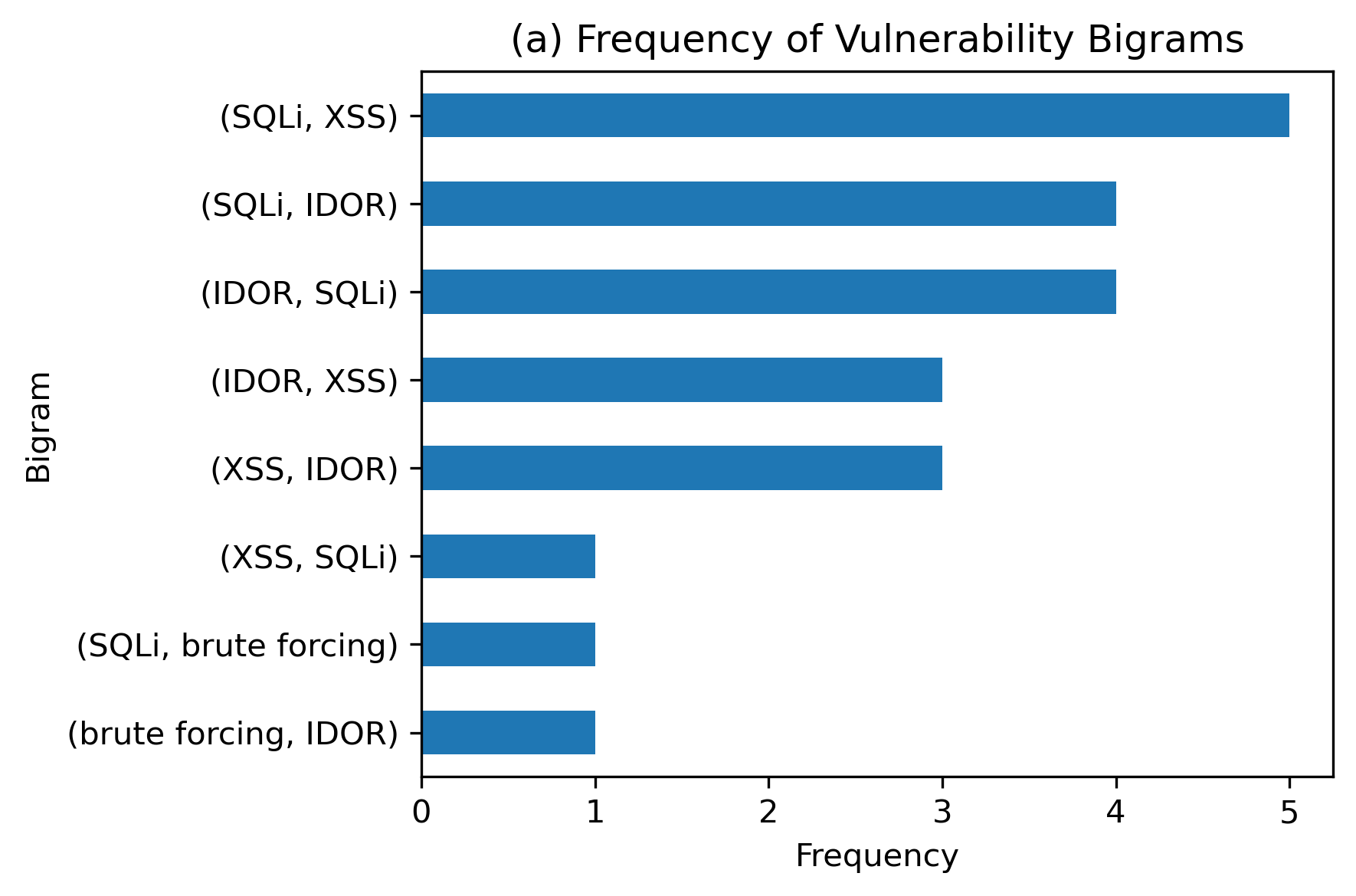}
        \caption{Frequency of cognitive patterns for different categories of vulnerabilities.}
        \label{fig:bigram_vuln}
\end{minipage}
\hfill
\begin{minipage}{.48\textwidth}
    \vspace{10ex}
    \centering
        \includegraphics[width=\textwidth]{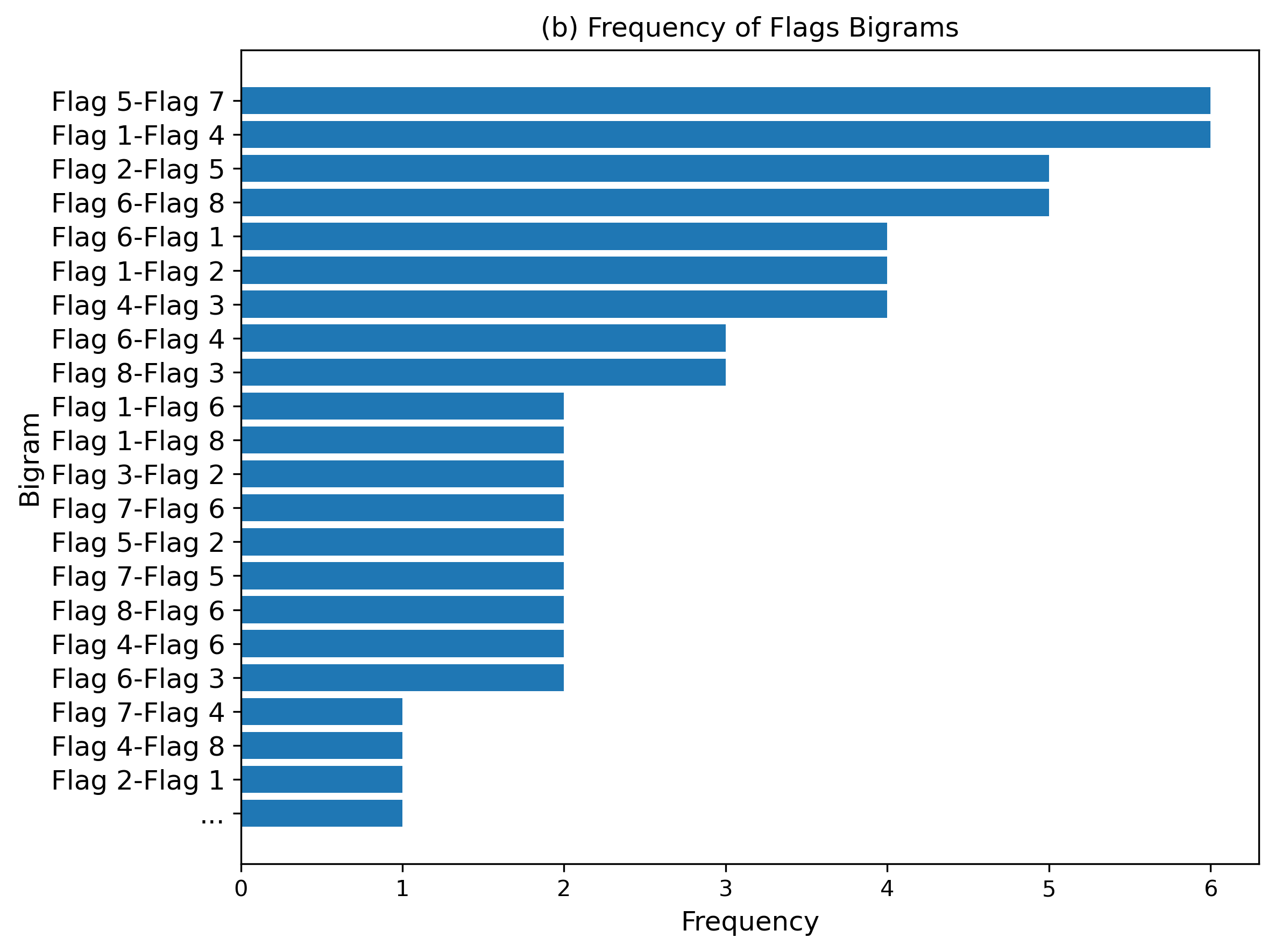}
        \caption{Frequency of cognitive patterns for different flags. SQL injection corresponds to flag1, flag2, flag4; IDOR corresponds to flag5, flag6, flag7, flag8; XSS corresponds to flag3.}
        \label{fig:bigram_flag}
\end{minipage}
\end{figure}

We organized the participants' actions into sequences representing the order in which they explored different vulnerabilities. Our analysis (Fig.~\ref{fig:bigram_vuln}) was then conducted to identify the frequency of bigrams in these sequences. 
The most frequent bigram observed is (SQL injection, XSS injection), occurring four times. This indicates that participants who identified an SQL injection vulnerability were likely to next search for an XSS injection vulnerability.

In addition to analyzing the sequences of vulnerabilities explored by participants, we also conducted a bigram analysis on the flags discovered by participants.
This analysis identified patterns in the types of vulnerabilities that participants tended to explore consecutively.
The resulting graph, as shown in Figure~\ref{fig:bigram_flag}, highlights the frequency of adjacent pairs of reported flags.
The top two switches is in the same category of vulnerability, suggesting that participants might have perceived certain types of vulnerabilities as more related or more likely to be found together, leading to a focused search strategy within the same type.
This insight provides valuable information on the decision-making processes and search strategies employed by participants during the CTF task, emphasizing the importance of understanding how perceived associations between vulnerabilities can influence search behavior.

To further understand the preferences and strategies of our participants, we conducted a ranking analysis on the sequence of successfully identified vulnerabilities. This analysis revealed patterns and preferences in how participants approached different types of vulnerabilities. 

\begin{figure}[tb]
    \centering
    \includegraphics[width=0.7\columnwidth]{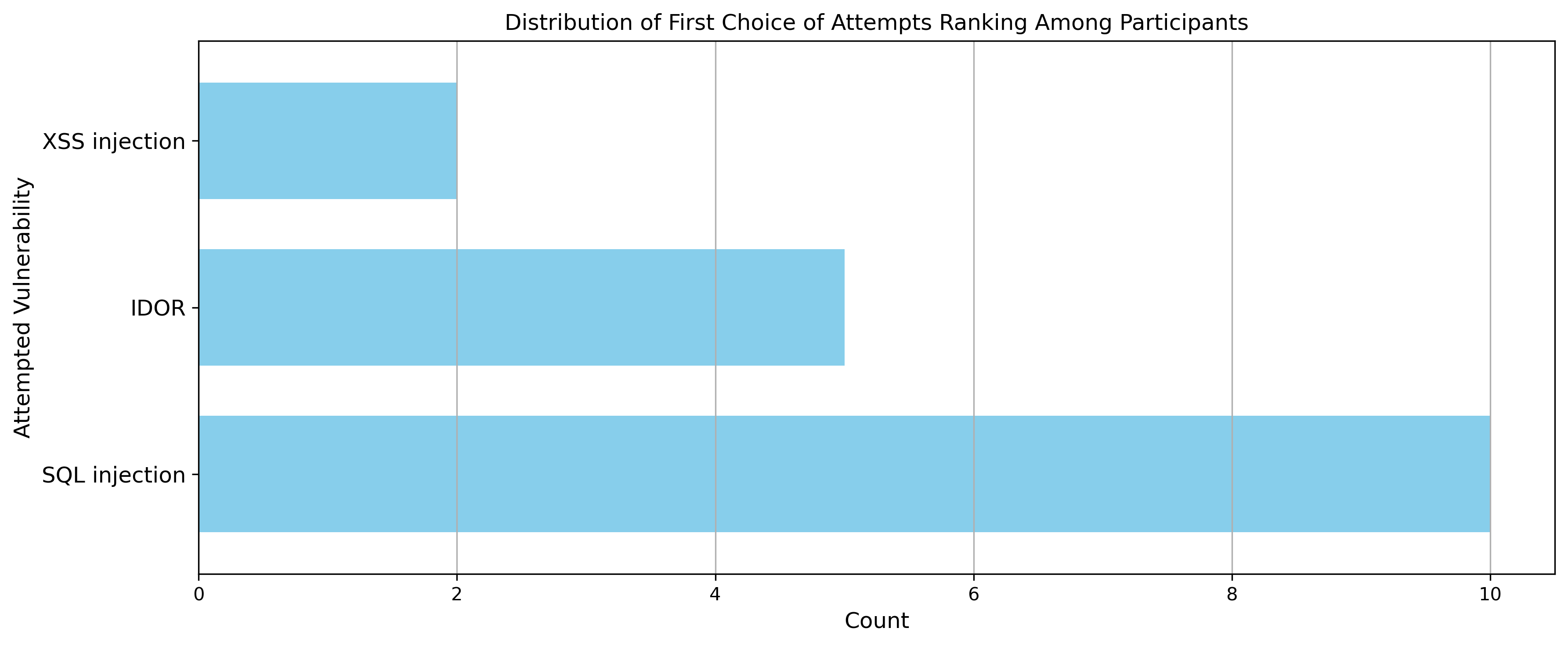}
    \caption{Distribution of participants' first-choice vulnerabilities for exploitation.}
    \label{fig:vuln_ranking}
\end{figure}

The results were visually represented in Figure~\ref{fig:vuln_ranking}, which highlights the initial choices made by participants: for the first successfully identified vulnerability, approximately 59\% of participants targeted SQL injection, 29\% on Insecure Direct Object References (IDOR), and 12\% focused on Cross-Site Scripting (XSS).
These findings suggest a potential prioritization or ease of identification associated with specific vulnerabilities.

\begin{tcolorbox}[colframe=black, colback=white, boxrule=0.5mm, arc=1mm, boxsep=1mm]
\textbf{Summary:} SQL injection is the first vulnerability most participants attempt, and they tend to continue exploring the same type of vulnerability throughout the exploration process.
\end{tcolorbox}

\section{Discussion and Limitations}

Our results indicate that participants who exhibited Satisfaction of Search (SoS) when exploring the system found significantly fewer vulnerabilities compared to those who did not exhibit this bias.
Additionally, participants who experienced Satisfaction of Search (SoS) reported a significant increase in pride and a decrease in nervousness.

We also explored the relationship between participants' self-efficacy beliefs and their actual performance in identifying vulnerabilities during Satisfaction of Search tasks.
To assess the connection between these variables, we employed both chi-square tests and Pearson correlation analyses.
The results from the chi-square test revealed no significant association between participants' reported levels of success and the number of vulnerabilities (flags) they successfully identified.
This finding was further corroborated by the Pearson correlation analysis, which also indicated a lack of significant correlation between self-efficacy and performance in the SoS task.
These results suggest that participants did not possess an accurate understanding of their performance.
This mismatch may lead participants to feel satisfied with their efforts, even when not all vulnerabilities have been uncovered.
Consequently, SoS can cause individuals to cease their search for additional threats following some initial discoveries,
potentially leaving significant vulnerabilities undiscovered.

Additionally, we investigated whether there is a relationship between SoS grouping and experience levels.
The results of a chi-square test of independence revealed no significant association between the SoS groups (satisfied vs.  not satisfied) and participants' experience levels.
This suggests that the influence of SoS is not dependent on the participants' experience levels, indicating that participants of all experience levels are largely equally susceptible to this cognitive bias.

\subsection{Implications}
In this section, we discuss the potential implications and limitations of our study. Our results show that Satisfaction of Search (SoS) can effectively influence participant performance.
In the context of cybersecurity education and CTF training, it is essential to address this bias explicitly. Educators should incorporate lessons on cognitive biases, such as Satisfaction of Search (SoS), into their curriculum to raise awareness among students. By understanding how these biases operate, students can be better equipped to recognize when they might be falling into such traps and take proactive measures to avoid them. Additionally, regular reminders and practical exercises that challenge students to continue their search even after finding initial vulnerabilities can help mitigate the influence of Satisfaction of Search (SoS), ultimately leading to more thorough and effective security assessments in both educational and competitive environments.

Furthermore, since CTF tasks can be used to observe attackers' behavior in real-world scenarios, the cognitive biases identified in this study could inform the design of honeypots. These honeypots would be effective in diverting penetration testers away from more valuable assets and providing early warnings about emerging threats~\cite{mokube2007honeypots}.
Integrating the SoS bias into honeypot design can amplify these functions.
By embedding multiple false vulnerabilities, honeypots can exploit the SoS bias, causing penetration testers to focus on non-critical aspects and miss genuine threats.
This distraction ~\cite{virtanen2022layer} can protect more valuable systems and data.
In addition, penetration testers who encounter convincing but ultimately insignificant data are likely to experience a false sense of accomplishment, leading them to prematurely cease their efforts.
The interactions with honeypots can further provide early warnings to administrators about ongoing attack attempts.
By using SoS to keep penetration testers engaged with the honeypot, administrators gain additional time to respond to and mitigate potential threats.
The prolonged engagement with honeypots, driven by SoS, allows for more detailed monitoring and analysis of penetration testers behavior, tactics, techniques, and procedures (TTPs).

\subsection{Limitations}
While our study provides valuable insights into the impact of cognitive biases on cybersecurity, it is important to acknowledge several limitations. 
One potential limitation of our study is its relatively small sample size. While we acknowledge this concern, it is important to note that similar recent research has been conducted with comparable sample sizes. For instance, many studies have drawn valuable conclusions from pools of no more than 15 participants~\cite{bertram2020trustworthiness, uwano2006analyzing,hu2024degpt} and some have even drawn conclusions from single digit pools~\cite{pollini2022leveraging, sharif2012eye}.
Such studies, including ours, provide preliminary insights and set
the stage for more extensive research in the future.

Secondly, the participants were recruited in a university setting from populations that had experience with Capture the Flag (CTF) competitions.
This background is not fully representative of real-world penetration testers, who may have varying motivations and levels of experience.
College students engaging in CTF challenges may exhibit skills and behaviors that differ from those of professional or malicious hackers targeting real systems.
However, previous research has used CTF competitions as proxies for real-world hacking campaigns~\cite{erola2021control, zennaro2023modelling}
In the case of real attacks, there is, to the best of our knowledge, no way of determining a participant's cognitive state or reason for halting their attack.
Thus, this is among the closest approximations available to the conditions of a real attack.

Lastly, the tasks designed for this study were based on CTF competition scenarios rather than real systems. The findings were collected based on a single web application developed for this experiment. While CTF tasks are useful for simulating certain aspects of cybersecurity challenges, they may not fully replicate the complexities and nuances of real-world systems that attackers encounter. Consequently, our findings might not extend to other, real-world applications or contexts outside of web security.


These limitations suggest that future work should aim to recruit a larger and more diverse sample of participants, including those with varied backgrounds and levels of experience in cybersecurity. Additionally, designing experiments that better mimic real-world systems and attack scenarios would provide a more accurate assessment of the impact of cognitive biases on attacker behavior.


\section{Conclusions}
Our study investigates the impact of cognitive biases, specifically Loss Aversion and Satisfaction of Search, on web application hacking attempts in CTF contexts. Through a controlled experiment with 17 CTF players acting as attackers, we systematically measured their performance and decision-making processes. We found that Satisfaction of Search significantly decreases an player's performance and alters their emotional state, while Loss Aversion does not notably impact their decision to continue tasks. Additionally, participants tended to repeatedly explore the same type of vulnerability. Our contributions include an IRB-approved experimental framework, and insights into the implications of these biases for cybersecurity education and defense strategies. This research highlights the potential of including cognitive biases in CTF training and leveraging cognitive biases to enhance defensive tactics against cyber attacks.

\section{Data Availability}
\label{availability}
All data and scripts are available at \url{https://osf.io/dy547/?view_only=d0489f50ae194c09be81a98fdfbcad54}.
This repository includes comprehensive documentation to facilitate replication and extension of our research.

\section{Acknowledgments}
The authors used ChatGPT for grammar check and text shortening when necessary.

\input{main.bbl}

\end{document}